\documentclass[12pt]{iopart}
\usepackage{graphicx}

\begin{document}

\title{On-chip SQUID measurements in the presence of high magnetic fields}

\author{Lei Chen}
\ead{leichen@magnet.fsu.edu}
\address{Department of Physics and The National High Magnetic Field Laboratory, Florida State University, Tallahassee, Florida 32310, USA}

\author{Wolfgang Wernsdorfer}
\address{Institut N\'{e}el, associ\'{e} \`{a} l'UJF, CNRS, BP 166, 38042 Grenoble Cedex 9, France}

\author{Christos Lampropoulos}
\address{Department of Chemistry, University of Florida, Gainesville, Florida 32611, USA}

\author{George Christou}
\address{Department of Chemistry, University of Florida, Gainesville, Florida 32611, USA}

\author{Irinel Chiorescu}
\address{Department of Physics and The National High Magnetic Field Laboratory, Florida State University, Tallahassee, Florida 32310, USA}

\begin{abstract}We report a low temperature measurement technique and magnetization data of a quantum molecular spin, by implementing an on-chip SQUID technique. This technique enables the SQUID magnetometery in high magnetic fields, up to 7 Tesla. The main challenges and the calibration process are detailed. The measurement protocol is used to observe quantum tunneling jumps of the $S=10$ molecular magnet, Mn$_{12}$-tBuAc. The effect of transverse field on the tunneling splitting for this molecular system is addressed as well.
\end{abstract}



\maketitle

Recent investigations on molecular magnets~\cite{1} or less complex systems containing diluted spins~\cite{2} have been motivated by their quantum mechanical properties and for future implementation in quantum computing~\cite{3}. Such studies require a large number of identical and independent magnetic cores (often with a large spin), to provide a natural amplification of the fundamental quantum mechanical phenomena of a single molecule to measurable signals. Remarkable quantum phenomena, such as the quantum tunneling of magnetization~\cite{4,5,6,7} and the Berry phase quantum interference ~\cite{8,9}, have been observed so far. Sensitive magnetization measurements in sweeping magnetic fields have been performed by means of superconducting quantum interference devices (SQUIDs)~\cite{1}, Hall probe magnetometers ~\cite{10} or Electron Paramagnetic Resonance (EPR)~\cite{11} spectroscopy. Such techniques are highly desirable to gather more information about effects such as the anisotropy-induced tunneling gap\cite{12}, spin states entanglements, phonon bottleneck effects ~\cite{13,14} and others.

SQUID devices are well established as highly sensitive magnetic flux detectors for low temperature applications. A commercial SQUID fabricated with Nb-AlO$_x$-Nb Josephson junctions is usually magnetically shielded and coupled to the sample through a pick-up coil~\cite{15}. This arrangement keeps the magnetic field from disturbing the SQUID. However, it also leads to a non-optimal coupling factor and excludes it from measuring samples of submicron size. A better coupling factor can be achieved by reducing the SQUID loop size and coupling to it directly, in the presence of a magnetic field. This represent an important advantage over Hall probes, since Josephson junctions can have widths of the order of the nanometer and have the sample on top or inside of it for optimal coupling. It is estimated that such SQUID implementation can detect, without extensive averaging, only few Bohr magnetons~\cite{16}. The planar tunneling junctions 
are not suitable because of their sensitivity to large magnetic field and relatively large dimensions.  Another type of Josephson junctions (nano-bridges)~\cite{17}  led to a promising design in which the SQUID is less sensitive to a field aligned along its plane, since the critical field value of a superconducting thin film can be much higher than that of the bulk in such a case~\cite{18}.  In this report, we present an improved on-chip SQUID measurement technique, in which a field up to 7~T can be aligned with high precision in the plane of a Nb SQUID. Thus, we have extended the observation of fundamental quantum phenomena by means of on-chip SQUID measurements into a higher magnetic field range, \textit{i.e.} to a larger family of molecular magnets. 

The technique allows real time (``on-the-fly") SQUID magnetization measurements on samples placed in the vicinity of the SQUID loop, to optimize the magnetic coupling. Both the sample and the SQUID detector are placed in the center of a 3-D vector magnet and thermally attached to the mixing chamber of a dilution refrigerator to ensure an efficient cooling.

\begin{figure}
\includegraphics[width=\columnwidth]{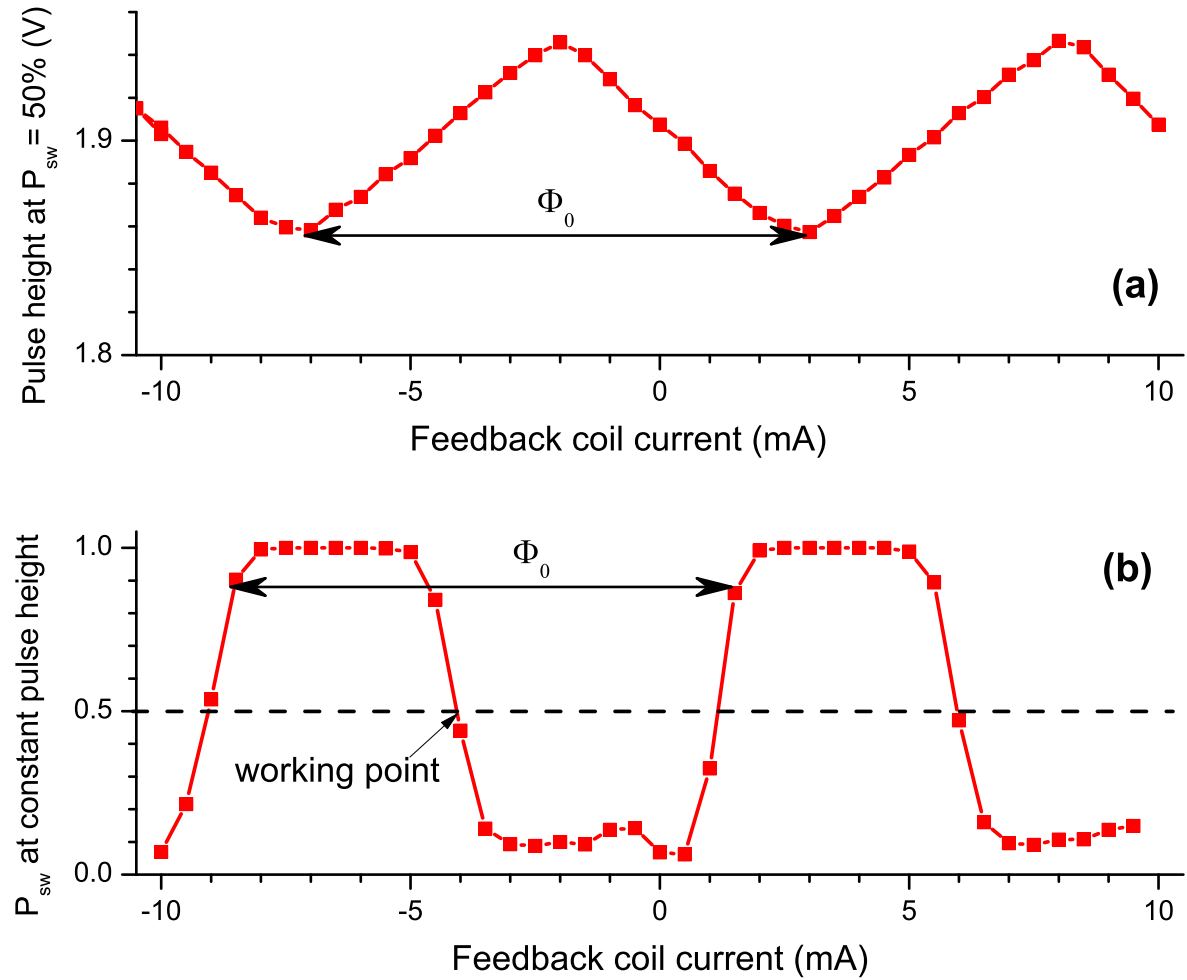}
\caption{(a) $I_{sw}$ modulation (shown as the height of voltage pulses applied on SQUID in series with a 15 k$\Omega$ resistor, such that $P_{sw}=50$\%) as a function of the field generated by the feedback coil (expressed in mA of coil current). (b) Same modulation as in (a), obtained by fixing the applied pulse height, such that $P_{sw}=50$\% for the designated working point, and ramping the coil current. The steep slopes leads to SQUID's high sensitivity.}
\label{fig1}
\end{figure}

The on-chip SQUID is made of a Nb film of only 5.5~nm thickness ensuring a superconducting state even for 7~T of in-plane magnetic field. The SQUID loop has a 2$\times$2 $\mu m^2$ area and contains two parallel nano-bridge Josephson junctions.  Therefore, the I-V characteristic of a SQUID is similar to a single Josephson junction with flux-dependent critical current and it can be explained with the resistively and capacitively shunted junction (RCSJ) model 
~\cite{18}. In the RCSJ model, the phase of a SQUID is described by the motion equation of a ball trapped in a washboard potential. An increase in the SQUID current tilts the washboard potential, and the phase ball has a probability to escape (and roll down along the tilted washboard) due to both thermal activation and quantum tunneling ~\cite{19}. The rolling of the phase ball generates a voltage across the SQUID and switches it from the superconducting state into the normal state. Therefore the switching probability $P_{sw}$ increases with the current amplitude. We define the switching current $I_{sw}$ as the current corresponding to $P_{sw}=50$\%~\cite{20}. In our experiment, current pulses are generated by a pulsed voltage source and an in-series resistor of 15~k$\Omega$ . The $I_{sw}$ current shows a typical periodic modulation as a function of the trapped magnetic flux in the SQUID loop (\Fref{fig1}(a)), with a period of $\Phi_0=h/2e$ ($h$ is Planck's constant and $e$ the electron charge). The modulation amplitude is 5\%, much smaller than typical values for SQUIDs with tunneling junctions but it can be used in fields up to 7~T.

Operating the SQUID in pulse mode, minimizes the heating generated by its switching into the normal state. The stochastic switching process is monitored by sending a series of current pulses of length $\sim 1\mu$s to the SQUID and monitoring the voltage across it. When the SQUID switches into the normal state (to a resistance of $\sim$4~k$\Omega$), voltage pulses are generated, amplified, and counted at the readout port. Statistically, the ratio of the counted output voltage pulses by the number of applied current pulses gives $P_{sw}$ for that current value. This switching probability is modulated by the magnetic flux, as shown in \Fref{fig1}(b).  The counting of switching events is expedited by operating the pulse mode at a high repetition frequency (here, 50~kHz). The mechanism used to perform an real time measurement will now be described. As shown in \Fref{fig1}(b), we define the middle point in the modulation curve as the "working point". During the measurement, a current pulse with a height corresponding to the working point is sent to the SQUID. It is important to note that the variation of the probability near the working point is very sharp, which is the key point behind the high flux sensitivity of a SQUID. 

To lock the SQUID at its working point and perform a feedback measurement, we place a small superconducting coil above the SQUID (see below \Fref{fig3}, insert) to actively compensate for flux changes caused by the magnetic sample. The feedback coil current is proportionally related to the change in the magnetic moment of the sample in an on-the-fly measurement.

Although the principle of the measurement is relatively simple, its practical realization is quite challenging.  The SQUID's thin film remains superconducting only when the high field is perfectly parallel to the plane. Even the smallest deviation from a precise alignment, can generate unwanted normal fields leading to vortices in the thin Nb film and altering the measurements. Large deviation simply switches the SQUID into the normal state. Also, the critical current of any superconducting material decreases with the increase of magnetic field, and so is the pulse height corresponding to the working point. The feedback mechanism only can be operated with a proper setting of the working point and therefore, the height of the current pulses corresponding to $P_{sw}=50$\% has to be carefully calibrated, as explained below. 

\begin{figure}
\includegraphics[width=\columnwidth]{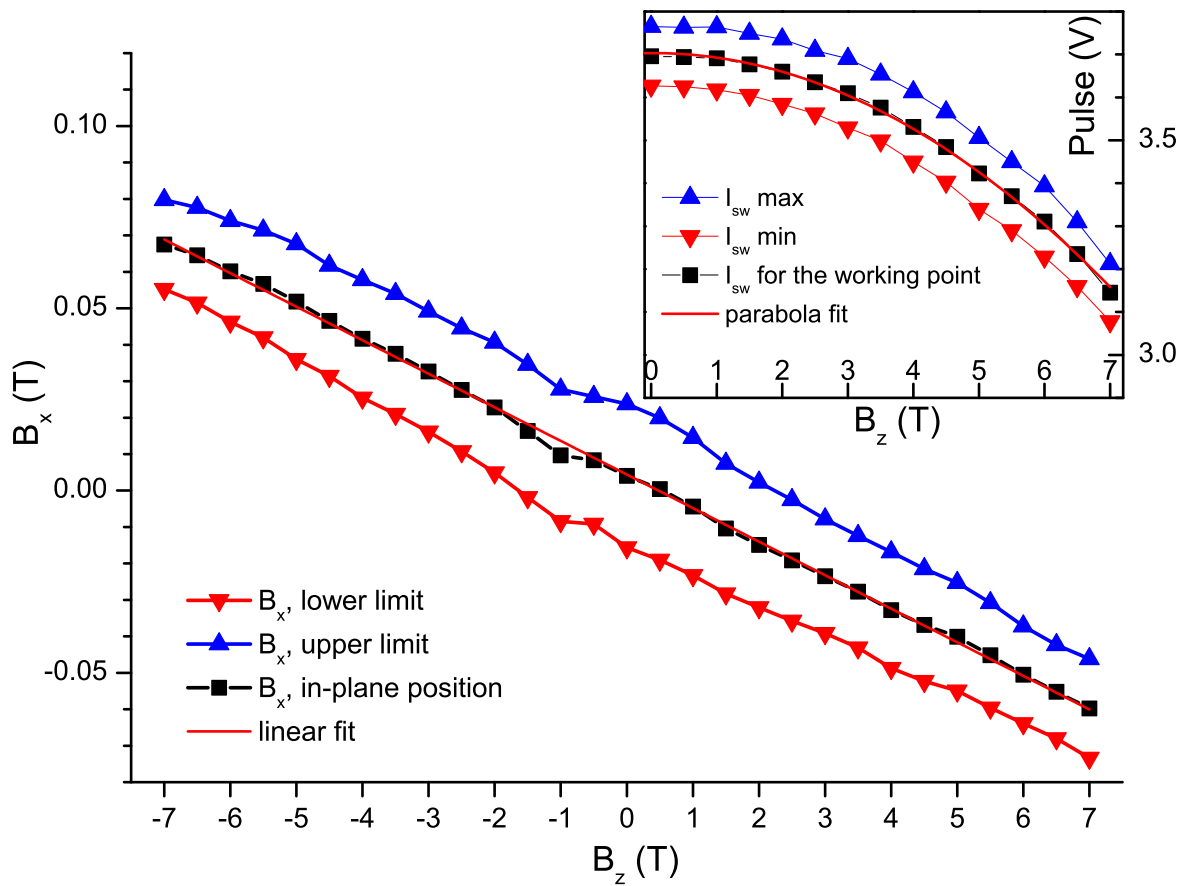}
\caption{The calibration process used to find the in-plane field as a combination of $\vec B_x$ and $\vec B_z$. The required $B_x$ (squares) is at the middle of the region (limited by triangles) where the SQUID is not switched into the normal state by the magnetic field. The linear fit leads to the misalignment angle between the holder and the coil system. (insert) Maxima and minima (triangles) of the $I_{sw}$ modulation as a function of $B_z$ (as in the main plot). The working point (squares) is chosen at half-distance and a parabolic fit calibrates the required decrease in the excitation pulse to keep the SQUID (in series with a 15 k$\Omega$ resistor) at $P_{sw}=$50\%.}
\label{fig2}
\end{figure}

\begin{figure}
\includegraphics[width=\columnwidth]{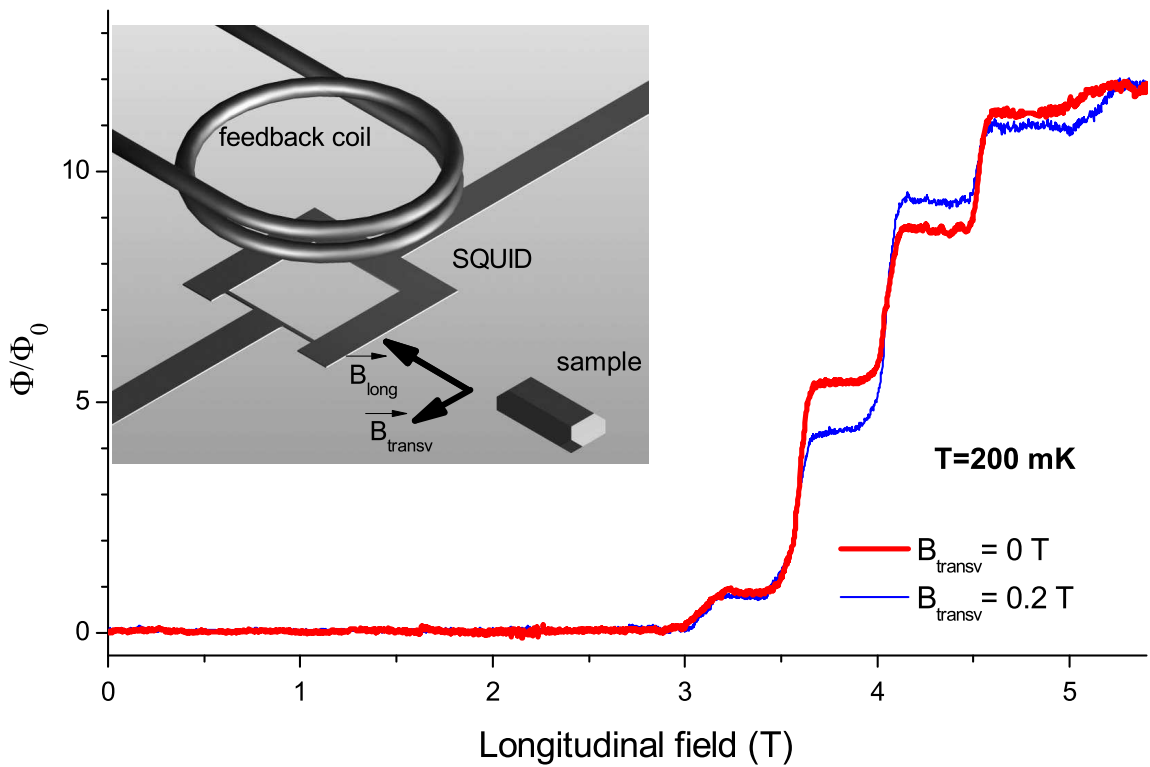}
\caption{Measured magnetization as a function of a longitudinal field applied along the easy axis of a Mn$_{12}$-tBuAc single crystal. The geometry is sketched in the insert: a single crystal is placed near an on-chip SQUID and under the feedback coil. The sample is first saturated at -5.5 T, and a real time measurement follows up to +5.5 T, here at a field rate of 1 mT/s. The two curves shows the differences in the quantum tunneling jumps - and therefore in the tunneling splittings- for two values of the transverse field, $B_{transv}$=0 T (thick line) and 0.2 T (thin line) respectively.}
\label{fig3}
\end{figure}

The precise alignment of the magnetic field in the SQUID plane is done by combining fields generated by a 3-axis magnet $\vec{B}=\vec B_x+\vec B_y+\vec B_z$. Initially, the SQUID plane is roughly aligned along $B_z$. Hence, we only need to scan $B_x$ to fine-tune the direction of the field in the SQUID plane. The SQUID can only sustain a small normal field ($\sim$15-20~mT), beyond which, the SQUID switches and stops working. In the calibration process, we scan $B_x$ (chosen roughly normal to the SQUID plane) for a series of given z fields, and record the field region in which the SQUID works (see \Fref{fig2}, triangles). The $B_x$ value in the middle of that region (shown with squares in \Fref{fig2}) combined with that given $B_z$ creates a total field in the SQUID plane.

During the field alignment described above, the maximum and minimum values of $P_{sw}$ modulation are recorded as a function of the magnetic field. These values are decreasing with field increase in a parabolic manner (see \Fref{fig2} insert), as a consequence of the critical current decrease, mentioned above. The working point is calibrated at the mid-distance between the two parabolas, so that the switching probability is kept constant at 50\% during SQUID operation. 

We demonstrate our first magnetization tunneling measurements with this high-field SQUID detection technique, on the Mn$_{12}$-tBuAc crystal, a molecular magnet requiring a high field (5.5 T) to reach magnetization saturation. A trace, starting from high negative fields (negative saturation) and up to $+$5.5 T, is shown in \Fref{fig3}. (thick line). Tunneling steps are clearly visible (see \cite{21} for a similar Hall-probe based measurement). Also, we were able to construct a constant transverse field normal to the crystal's magnetic easy axis, while keeping the total field perfectly aligned in the SQUID plane during the sweep of the field in the longitudinal direction. The magnetization curves of the sample show differences in the size of the tunneling steps (\Fref{fig3}, thin line). This indicates that the transverse magnetic field modifies the size of the tunneling gaps of this molecular magnet, similarly to a Berry-phase process (as in \cite{8}).

We have implemented the described feedback mechanism, simultaneously to two identical, neighboring SQUIDs. In this arrangement, the two SQUIDs sense similar external background field. If the studied sample is placed between the two SQUIDs, the magnetic flux lines generated by it are penetrating the SQUID loops in opposite directions. Therefore, a simple subtraction of two measurements can eliminate the uniform background and double the magnetic signal. Individual runs on each (empty) SQUID show a certain wavy background signal. Following their subtraction, which can be done in real time during a field sweep, the background is significantly flattened, suggesting that such a differential readout can further improve sensitive SQUID-based magnetic detection.

In conclusion, our on-chip SQUID detection has extended the possibility of performing magnetization studies on quantum spins into the high field regime (up to 7 T). This versatile technique will allow the exploration of the quantum properties of more types of magnetic molecules. During its operation, the sweeping magnetic field has to be perfectly aligned in the SQUID plane and the excitation current pulses have to be properly calibrated with field. The magnetic detection is using the feedback information from a superconducting coil fixing the working point of the SQUID. We have also succesfully tested a differential two-SQUID method which significantly reduces unwanted background signals and thus improves the SQUID capability. 

We acknowledge support from the NSF Cooperative Agreement Grant No. DMR-0654118, the NSF-CAREER grant No. DMR-0645408 and the Alfred P. Sloan Foundation. WW thanks the financial support from the ANR-PNANO projects MolNanoSpin ANR-08-NANO-002 and the ERC Advanced Grant MolNanoSpin 226558.

\noappendix

\end{document}